\documentclass[%
 reprint,
 footinbib,
 showkeys,
 amsmath,amssymb,
 aps,
]{revtex4-2}

\usepackage{graphicx}
\usepackage{epstopdf,subfig,epsfig}
\usepackage{floatrow}

\usepackage{dcolumn}
\usepackage{bm}

\usepackage{enumitem}
\usepackage{etoolbox}
\usepackage{textgreek}
\usepackage{xcolor}

\renewcommand{\d}{\mathrm{d}}

\newcommand{\const}{\mathrm{const}}

\newcommand{\be}{\begin{equation}}
\newcommand{\ee}{\end{equation}}
\newcommand{\bea}{\begin{eqnarray}}
\newcommand{\eea}{\end{eqnarray}}
\newcommand{\ba}{\begin{equation}\begin{aligned}}
\newcommand{\ea}{\end{aligned}\end{equation}}

\newcommand{\edit}[1]{{\color{black}{#1}}}

\begin{document}


\title{Physics of a granular pile}

\author{R. Krechetnikov}
\affiliation{Mechanical and Civil Engineering, California Institute of Technology, Pasadena, CA 91125, USA}
\email{krechet@ualberta.ca}
\altaffiliation[on leave from ]{Mathematics Department, University of Alberta, Edmonton, AB, T6G 2G1, Canada}
\author{A. Zelnikov}
\affiliation{Physics Department, University of Alberta, Edmonton, Alberta, T6G 2E1, Canada}


\begin{abstract}
From the onset of the subject, granular media have been defying the toolkit of statistical mechanics thus hindering our understanding of their thermodynamical and rheological properties and making this state of matter one of the key remaining mysteries in science. In the present work, we offer a resolution to this problem in the case of static granular media by considering a collective behavior of 2D identical balls forming a granular pile in the gravity field, which allows us to develop its thermodynamics and rheology from the first principles. Besides the uncertainty due to rough substrate on which the pile is built, we uncover another one due to ambiguities occurring in the positions of some interior balls. Both are responsible for the \textit{thermodynamic description} of the granular pile, which proves to be anything but ordinary. In particular, we show that a pile is characterized by three temperatures: one is infinite, the other is negative, and the third is of higher-order. Also, considering the fields of ball displacements $\boldsymbol{\xi}$ and the normal force deviations $\delta\boldsymbol{N}$ from the ideal isosceles triangular structure of a regular pile reveals the hyperbolic nature of the $\boldsymbol{\xi}$-field and ability of the $\delta\boldsymbol{N}$-field to change the characteristic type from hyperbolic to elliptic. The latter property not only explains the origin of \textit{force chains} and provides an adequate description of the \textit{rheology} of static granular media, but turned out to be instrumental for understanding the granular media thermodynamics. Our model should provide a basis for further grasping of granular media properties, in general.
\end{abstract}



\maketitle


\section{Introduction and setting} \label{sec:Intro}

\vspace{-0.25cm}

Despite that literature on granular matter (GM) is as extensive as its presence in nature and life \cite{Duran:2000,Tahmasebi:2023}, it remains a largely unresolved challenge in science \cite{Kennedy:2005}. Having significant economic and industrial implications in civil, chemical, metallurgical, pharmaceutical, and food engineering as well as being a crucial component in geo- and astrophysics make the gap in understanding of their fundamental rheological and thermodynamical properties even more glaring. At the same time, humans' fascination with many properties of granular media distinguishing them from liquids and solids has been never-ending: moving desert dunes \cite{Bagnold:1941}, trapping due to liquefaction of quicksands \cite{Khaldoun:2005,Khaldoun:2006}, expanding in volume as GM is sheared under the foot of a person walking on beach resulting in drying up wet sand \cite{Reynolds:1885}, size segregating with larger particles ending up on top \cite{Kudrolli:2007}, arching when stresses distribute with stiffer components of the system attracting more load \cite{Nadukuru:2012} and hence leading to \textit{force-chains} \cite{Majmudar:2005,Krim:2009}, and so forth. Despite these and many other marvels, wealth of experimental observations, and abundance of phenomenological models, there is still lack of the very basic fundamental understanding when the above properties would be predicted from a model deduced \textit{ab initio}.

\edit{Without pretending to the completeness of literature review, we provide the key motivational points for the present study. Search for a continuous \textit{rheological model} of GM has usually been guided by the presumption that GM must be placed in between liquids and solids and hence stresses are sought as a function of strain both in static \cite{Babic:1997,Ball:2002,Blumenfeld:2004} and flowing \cite{Johnson:1987,Babic:1997} GM, to name a few. At some point, the limited of success of GM rheological models motivated challenging the validity of continuum postulates \cite{Rycroft:2009}. The existence of force-chains, which make GM inherently \textit{inhomogeneous}, led some authors \cite{Henann:2014,Jop:2015,Kamrin:2024} to abandon a local continuum description and search for non-local models, in the sense of involving higher gradients of stress or strain.}

\begin{figure}[h!]
	\setlength{\labelsep}{3.0mm}
	\centering
    \sidesubfloat[]{\epsfig{figure=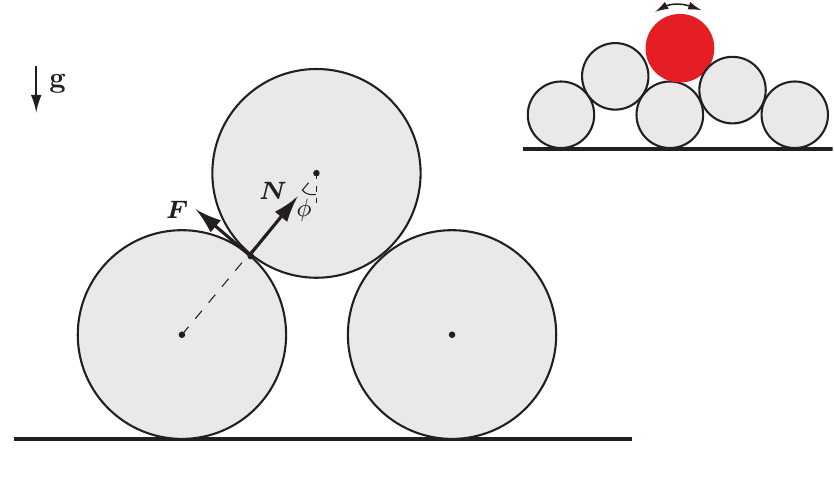,height=1.65in}\label{fig:olympiad}} \setlength{\labelsep}{1.0mm} \\[0.5cm]
	\sidesubfloat[]{\epsfig{figure=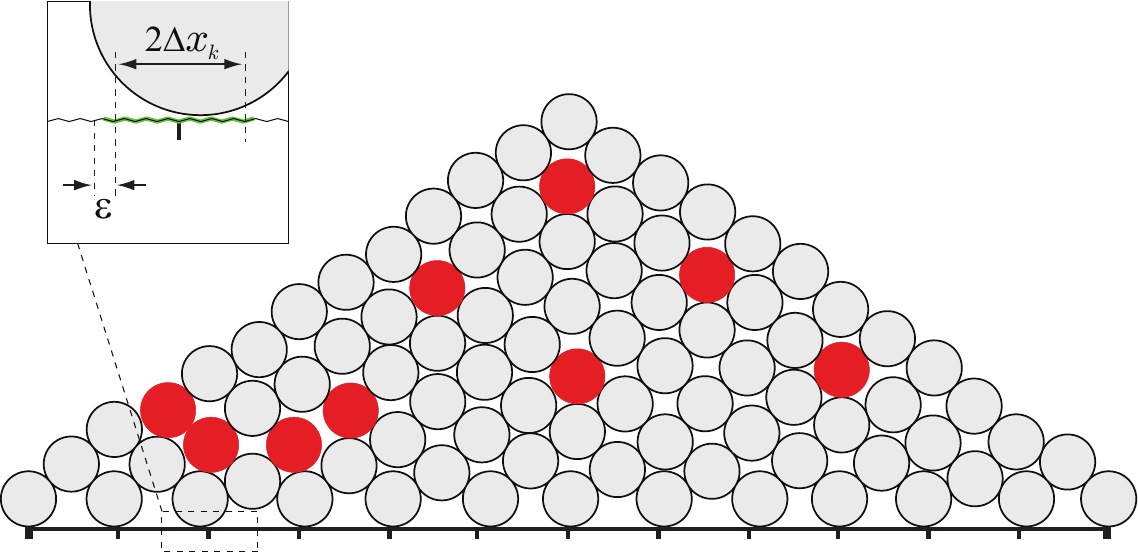,height=1.45in}\label{fig:setting}}
\caption{\setlength{\rightskip}{0pt} \setlength{\leftskip}{0pt}Setting: (a) Three-ball problem set in the gravity field $\mathbf{g}$ with $\boldsymbol{N}$ and $\boldsymbol{F}$ denoting the normal and friction forces, respectively, acting on the upper ball; inset with six balls shows the emergence of a fork \edit{-- ambiguity in the (red) ball position}. (b) A pile with $n$ balls in the first layer, the end ones of which are fixed and the interior $n-2$ ones are displaceable; the balls are of unit radius $r=1$, mass $m$, and nearly rigid \edit{as discussed in the main text}; \edit{red balls mark the location of forks}. The first layer is sitting on a rough substrate with groove size $\varepsilon$ responsible for the discreteness of the ball positions; this layer is characterized by the average gap \edit{$0 \le \Delta \ell < 2$} between balls, and the maximum allowed amplitude $\Delta x_{k} = \frac{k}{k_{\mathrm{max}}} \Delta x_{\mathrm{max}} \le \Delta x_{\mathrm{max}} = \min{\left(\frac{\Delta \ell}{2},1-\frac{\Delta \ell}{2}\right)}$, $k=0,\ldots,k_{\mathrm{max}}$ of interior ball displacements relative to a regular (isosceles triangular) pile, $k=0$ (marked with ticks); $\Delta x_{k}$ is a multiple of $\varepsilon$; the probability distribution of ball positions is homogeneous on the interval $[-\Delta x_{k},\Delta x_{k}]$ centered at the regular pile ball position.}
\end{figure}
On the \textit{thermodynamics} side of the story, Edwards \cite{Edwards:1994} suggested to consider the number of ways one can assemble $\mathcal{N}$ grains to fill the actual volume $V$ taken up by the granular medium, the logarithm of which is the entropy $S_{\mathrm{Ed}}(V, \mathcal{N})$. The latter, in analogy to temperature in the standard thermodynamics, gives $X = \left(\partial V / \partial S_{\mathrm{Ed}}\right)_{\mathcal{N}}$ named as the compactivity measuring the ``fluffiness'' of the medium. Edwards also assumed that when $\mathcal{N}$ grains occupy a volume $V$ they do so in such a way that all configurations are equally weighted with the Boltzmann-type factor $e^{-V/X}$ being equivalent to the ergodic hypothesis in thermal physics \cite{Reif:1965,Callen:1985}. However, the granular systems proved not to exhibit ergodicity \cite{Paillusson:2012,Blumenfeld:2016}. As a result, Blumenfeld et al. \cite{Blumenfeld:2016} proposed an alternative formulation based on a connectivity function that depends on all the structural degrees of freedom, but this formulation has been criticized as well \cite{McNamara:2016}, in particular because the connectivity has no obvious macroscopic meaning, and hence the question \edit{of adequate temperature-like variables in the static \cite{Baldassarri:2005,Henkes:2007,Bi:2015,Jiang:2017,Ball:2019,Bililign:2019} and dynamic \cite{Baldassarri:2005,Bi:2015,Jiang:2017,Maranic:2021}} GM remains open despite recurring attempts.

In this work we \edit{focus on the static case of GM and develop a bottom-up approach to its thermodynamics and rheology. Starting} with the 2D problem of three balls (disks) in the gravity field $\mathbf{g}$, cf. fig.~\ref{fig:olympiad}, offered in 1986 at a theoretical mechanics olympiad \cite{Popov:2006}, we learn that the minimum static friction coefficient $\mu_{\mathrm{s}}(\phi) \in \left[2-\sqrt{3},1\right)$ is required for stability of the pile with the opening angle $\phi \in \left[\frac{\pi}{6},\frac{\pi}{2}\right)$, \edit{cf. Appendix~\ref{subsec:3-balls}}. While in the $3$-ball problem the friction force $\boldsymbol{F}$ and hence \edit{its ratio to the normal $\boldsymbol{N}$ force, i.e. a local friction coefficient $\mu = |\boldsymbol{F}|/|\boldsymbol{N}|$,} turn out to be unique, in general it can assume a range of values as permitted by stability of a GM system; \edit{note that thereby defined $\mu \le \mu_{\mathrm{s}}$ should not be confused with the static friction coefficient $\mu_{\mathrm{s}}$ at which the relative motion occurs \cite{Persson:1998}}. Indeed, extending to six balls as in the inset of fig.~\ref{fig:olympiad}, we immediately encounter two phenomena: (i) a possibility for the top ball to assume two possible positions when leaned against either the left or the right surrounding balls -- what we call a \textit{fork} -- and (ii) the friction forces acting on that ball can assume any magnitude in the interval $F \in[-\mu |\boldsymbol{N}|,+\mu |\boldsymbol{N}|]$. While the latter \textit{indeterminacy} of friction phenomena is usually deemed as a complication \cite{Andreotti:2013}, in the present work we use it to our advantage to \textit{decouple} the \textit{geometric} and \textit{force} problems, i.e. in our case mechanical stability does not have an effect on packing as opposed to dynamic GM problems \cite{Heitkam:2012}.

Namely, first we generate a random distribution of $n$ balls in the first layer \edit{placed on a substrate undulated with the (roughness) period $\varepsilon$, responsible for the discreteness of ball positions as in fig.~\ref{fig:setting}}, upon which a static (glassy) pile of $\mathcal{N} = \frac{1}{2} n (n-1)$ balls is built \edit{by placing balls layer by layer} as in fig.~\ref{fig:setting} thus constituting the geometric problem; \edit{in the process, forks are encountered as marked by red balls in fig.~\ref{fig:setting}. The bottom layer is characterized by the average gap $0 \le \Delta \ell < 2$. The allowed amplitude (dispersion) $\Delta x_{k} = \frac{k}{k_{\mathrm{max}}} \Delta x_{\mathrm{max}} \le \Delta x_{\mathrm{max}} = \min{\left(\frac{\Delta \ell}{2},1-\frac{\Delta \ell}{2}\right)}$, $k=0,\ldots,k_{\mathrm{max}}$, of interior ball displacements (in the first layer) relative to the regularly spaced ball positions is chosen such that the balls from the second layer do not fall through to the bottom layer, which guarantees that we work with triangular piles having the same total number of balls $\mathcal{N}$. The bottom layer balls probability distribution is homogeneous in each interval $[-\Delta x_{k},\Delta x_{k}]$, divided into $2 \Delta x_{k}/\varepsilon$ groves the number of which can be as large as we want for $\varepsilon \rightarrow 0$. Then we} extend the $3$- to $\mathcal{N}$-ball analysis (see Appendix~\ref{subsec:linear-programming}), which involves balancing torques and force projections, to find a solution to the static force problem subject to minimization of the elastic energy functional $U = \sum_{i,j}{\left(|\boldsymbol{N}_{i}^{j}|^{2} + |\boldsymbol{F}_{i}^{j}|^{2}\right)}$ with the summation \edit{over each contact point $(i,j)$ between all balls in the pile, cf. fig.~\ref{fig:force-notations}}. \edit{This form of elastic energy follows from the assumption of balls being nearly rigid, so that all elastic deformations are concentrated near the point of the net force $\left(|\boldsymbol{N}_{i}^{j}|^{2} + |\boldsymbol{F}_{i}^{j}|^{2}\right)^{1/2}$ application and therefore the stated form of $U$ follows from the solution of the problem of a half-plane deformation under a point load \cite{Landau:1986}. This quadratic energy functional is subject to linear equality constraints -- equations for balance of forces and torques -- and inequality constraints -- non-negativity of normal forces -- as per Appendix \ref{subsec:linear-programming}, the problem reduces to the standard quadratic optimization problem, the solution of which is guaranteed by the Karush-Kuhn-Tucker theorem \cite{Luenberger:1969}.} If such a local minimum solution exists (which is physically guaranteed by autotuning local friction coefficients as in \edit{dynamical systems \cite{Kumar:2015}, where ``they access a large range of effective friction coefficients that allows self-tuning of the system to adjust its response to changing environments''}), the pile is mechanically stable in the Lyapunov sense \cite{Krechetnikov:2007,Heitkam:2012} and hence physically realizable. We also treat GM as \textit{athermal}, which can be justified if a pile is always in thermal equilibrium with the environment. It is the gravity field and ability to adjust the friction forces without affecting the geometric fabric of GM that play the role of an external energy reservoir in our case. Namely, once we constructed a geometric fabric of our GM pile, turned on the gravity, all the energy of the resulting change in the pile center of mass is converted into the energy of elastic deformations. On top of that, by variation of the friction forces, we may add the corresponding extra elastic deformations. Hence, all the energy we have to deal with is elastic as per the above introduced functional $U$. The presented problem formulation captures the GM key ingredients, namely it is formed in a natural setting due to cohesion mediated by gravity \footnote{as otherwise van der Waals forces between GM elements are negligible compared to the ball weights and hence normal as well as friction forces}, which brings about the pile intrinsic \textit{inhomogeneity} and \textit{non-isotropy} as $\mathbf{g}$ sets a preferred direction and ``hydrostatics'' \footnote{Note that we use the term ``hydrostatic'' even though our system is not fluid, but the usage will be justified in due course.}. Nevertheless, as we will see, this does not prevent us from formulating a well-defined thermodynamic picture and is, in fact, essential for the proper understanding of GM rheology.

\vspace{-0.35cm}


\section{Thermodynamics} \label{sec:thermo}

\vspace{-0.35cm}

\begin{figure}[h!]
	\setlength{\labelsep}{-4.0mm}
	\centering
    \sidesubfloat[]{\epsfig{figure=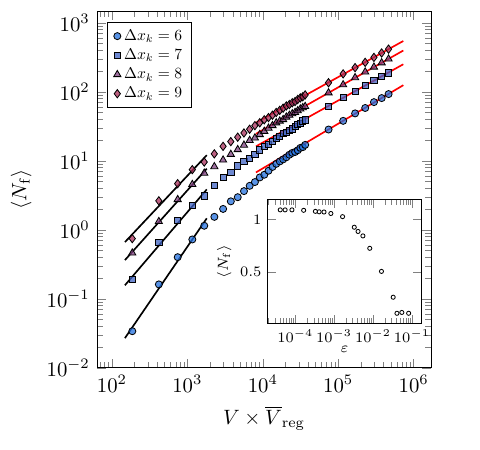,height=3.25in}\label{fig:extensivity}} \\
	\sidesubfloat[]{\epsfig{figure=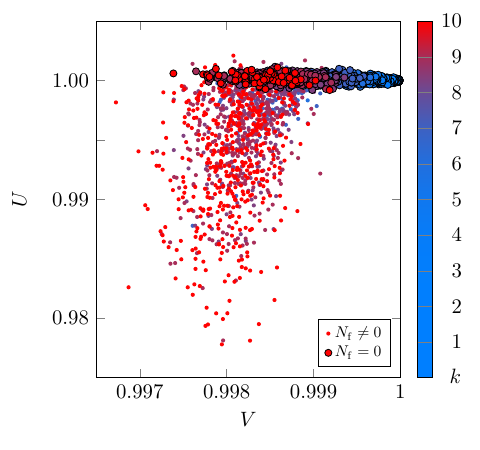,height=3.25in}\label{fig:energy-area}}
\caption{\setlength{\rightskip}{0pt} \setlength{\leftskip}{0pt} Exploratory observations: (a) the \edit{mean number of forks $\langle N_{\mathrm{f}} \rangle$}, \edit{obtained by averaging over all realizations for a given $\Delta x_{k}$,} vs dimensional pile volume $V \times \overline{V}_{\mathrm{reg}}$ for several values of $\Delta x_{k}$, with the limiting average slope for large values of volume being \edit{$\langle N_{\mathrm{f}} \rangle = (V \times \overline{V}_{\mathrm{reg}})^{0.620 \pm 0.004}$}, the range $n=11-501$ and $\Delta \ell = 1.3$; volume is measured as the interior area of the polygonal chain obtained by connecting the ball centers along the entire pile perimeter; the inset indicates that the mean number of forks $\langle N_{\mathrm{f}} \rangle$, \edit{obtained by averaging over all realizations for $\Delta x_{\mathrm{max}}$,} stabilizes with refining the roughness size $\varepsilon$ in the case $n=25$; (b) the pile elastic energy $U$ vs $V$ for $n=25$ and $\Delta \ell = 1.3$; both $U$ and $V$ are scaled w.r.t. the corresponding values of a regular pile.}
\end{figure}

An \textit{ensemble} of random realizations (microstates) of piles in mechanical equilibrium with the fixed number of balls $\mathcal{N}$, as defined in \S \ref{sec:Intro}, allows us to stay within the realm of equilibrium thermodynamics \edit{and to develop thermodynamical description by averaging over this ensemble or its subsets, e.g. corresponding to a fixed number of forks, as required by the underlying theory}. \edit{Whereas mathematically our balls are identical, they are distinguishable being marked by the position $(i,j)$, so there is no entropy reduction associated with the exchange of the balls which would otherwise appear in a collection of identical and indistinguishable particles such as in quantum thermodynamics.} Hence, our system is \textit{classical} with \textit{two sources} of uncertainty -- random ball distribution on a rough substrate as in fig.~\ref{fig:setting}, which we call collectively as \textit{roughness}, and \textit{forks} -- resulting in the corresponding entropies analogous to the considered in the literature packing \cite{Woodcock:1997,Logan:2022} and configurational \cite{Sutton:2020} entropies, which we take to be dimensionless \cite{Leff:1999} and thus temperature has units of energy. For the allowed interval of ball positions $\Delta x_{k} \ge \varepsilon$, cf. fig.~\ref{fig:energy-area}, we have $N_{\mathrm{r}} = (1 + 2 \Delta x_{k}/\varepsilon)^{n-2}$ possible configurations in the first layer thus furnishing an informational \textit{roughness entropy} $S_{\mathrm{r}} = \ln{N_{\mathrm{r}}}$ with $S_{\mathrm{r}} = 0$ for a regular pile as required; this definition of entropy reflects the amount of information we do not know about the system \cite{Jaynes:1957}. Appearance of forks provides yet another independent source of uncertainty as a ball in the fork may assume either of the two possible positions. In contrast to the roughness entropy, the entropy of forks $S_{\mathrm{f}} = N_{\mathrm{f}} \ln{2}$ -- the formula for which follows by construction of a pile from bottom up \edit{\footnote{When a fork is encountered, we chose randomly either left or right ball position resulting into two realizations or more, if new forks are encountered in upper layers. This accounts for the fork entropy $S_f = \ln{2^{N_f}}$ for a given pile realization with $N_f$ forks. Alternatively, validity of this formula can be demonstrated empirically by constructing many pile realizations for the same bottom layer distribution and calculating the probability of each realization with $N_{f}$ forks, which proves to be $2^{-N_{f}}$.}} -- is not readily computed since the number of forks $N_{\mathrm{f}}$ is not known a priori and, in general, dependent on the roughness size $\varepsilon$, cf. inset in fig.~\ref{fig:extensivity}: in the present work we will focus on the case when $\varepsilon$ is sufficiently small so that $N_{\mathrm{f}}$ is independent of it. Most importantly, fig.~\ref{fig:extensivity} indicates that entropy of forks is not extensive, i.e. not proportional to the pile volume \footnote{or, equivalently, the total number balls $\mathcal{N}$ since in our case $V \sim n^{2.0355 \pm 0.0039}$ over the range $n=11-501$}, the reason for which will become clear in the context of the rheology discussion (\S \ref{sec:rheol}); \edit{the pile volume is measured as the interior area of the polygonal chain obtained by connecting the ball centers along the entire pile perimeter}. While non-extensivity is not unusual in other physical systems \cite{Maddox:1993,Tsallis:2005,Tsallis:2009,Becattini:2019} such as black holes, it does not allow us to use a more convenient for computations integral form of the fundamental thermodynamic relation, but, instead, its differential form
\begin{align}
\label{relation:thermodynamic-fundamental}
\d U = T_{\mathrm{r}} \d S_{\mathrm{r}} + T_{\mathrm{f}} \d S_{\mathrm{f}} - P \d V \equiv T_{\mathrm{f}} \d S_{\mathrm{f}} - P_{\mathrm{tot}} \d V,
\end{align}
\edit{where for \textit{universality} of the results representation, we normalized the volume by that of a regular pile $\overline{V}_{\mathrm{reg}}(n,\Delta \ell) = \frac{1}{4} (n-1)^{2} (2 + \Delta \ell) \sqrt{16 - (2 + \Delta \ell)^{2}}$ with the same $n$ and $\Delta \ell$ \footnote{Notably, the volume of irregular piles follows approximately the same curve with the larger deviations at $\Delta \ell \sim 1$.}, the energy by that of a regular pile $\overline{U}_{\mathrm{reg}}(n,\Delta \ell)$, so that temperatures $T_{\mathrm{r}}$, $T_{\mathrm{f}}$ are also normalized by $\overline{U}_{\mathrm{reg}}$, whereas pressure $P$ by $\overline{U}_{\mathrm{reg}}/\overline{V}_{\mathrm{reg}}$. In \eqref{relation:thermodynamic-fundamental} we} did not include the chemical potential contribution as the number of balls $\mathcal{N}$ is fixed and introduced \edit{total pressure} $P_{\mathrm{tot}}$ on the rhs based on the a posteriori discovered equation of state relating \edit{the original pressure} $P$ and $T_{\mathrm{r}}$ in the considered limit of small $\varepsilon$. Since $\Delta x_{k}$ is our probability parameter controlling the degree of deviation from a regular pile, all physical variables such as energy $U$, etc. will be computed as its functions; hence, for calculating differentials we will vary $\Delta x_{k} \rightarrow \Delta x_{k} + \delta x$, which will provide variations of the corresponding variables, e.g. $\delta U = U(\Delta x_{k} + \delta x) - U(\Delta x_{k})$.

Whereas fig.~\ref{fig:extensivity} demarks the near-extensive behavior of the fork entropy $S_{\mathrm{f}}$ for small volumes \footnote{for the lowest shown value of $\Delta x_{k}=6$ the fluctuations are smaller and hence appearance of forks is more rare -- and therefore, in general, non-extensive} and non-extensive for large volumes, plot $U(V)$ in fig.~\ref{fig:energy-area}, from which all exploratory thermodynamic studies usually start \cite{Leff:2021}, clearly tells us that the appearance of forks is responsible for some fundamentally important underlying physical processes as the energy relaxation from its no-fork values becomes substantial. Also, fig.~\ref{fig:energy-area} alone indicates that volume $V$ (and hence compactivity $X$) is not the right parameter to deal with in GM as there is no one-to-one correspondence with other properties; rather, it is the ball arrangement and force distribution which control the system behavior. With such an \textit{enigma}, we will embrace in understanding thermodynamical and rheological properties of an ordinary pile of balls.

\begin{figure*}
	\setlength{\labelsep}{-4.0mm}
	\centering
    \sidesubfloat[]{\epsfig{figure=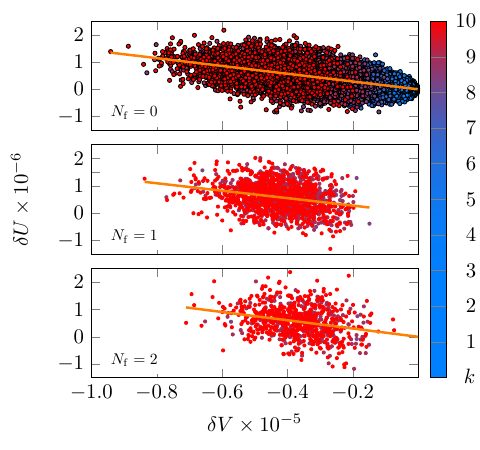,height=3.25in}\label{fig:denergy-darea-3}} \quad
	\sidesubfloat[]{\epsfig{figure=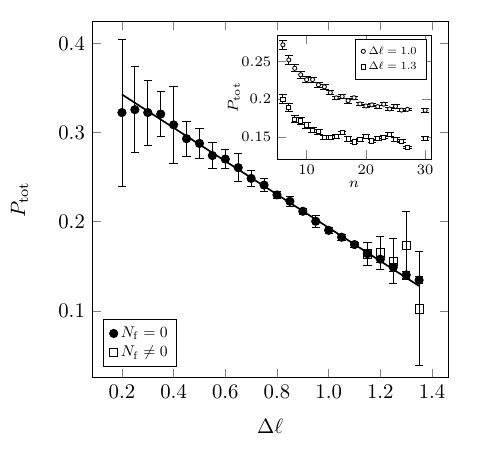,height=3.25in}\label{fig:pressure}} \\
    \sidesubfloat[]{\epsfig{figure=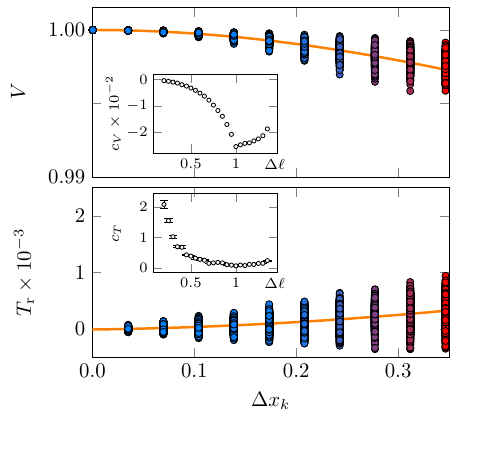,height=3.25in}\label{fig:roughness}} \quad
	\sidesubfloat[]{\epsfig{figure=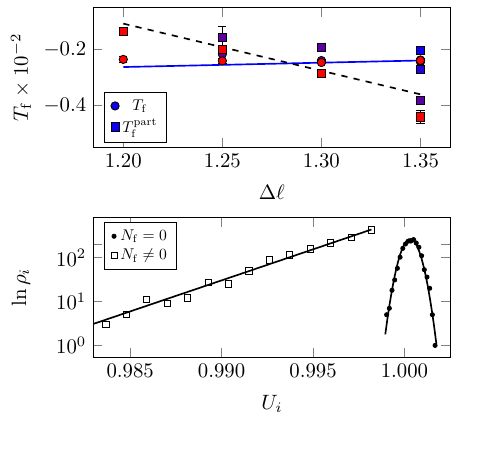,height=3.25in}\label{fig:temperature}} \\
\caption{\setlength{\rightskip}{0pt} \setlength{\leftskip}{0pt} Thermodynamics: (a) \edit{representative data for $\delta U$ vs $\delta V$} for $\Delta \ell=1.3$, $n=25$, and three values of the number of forks $N_{\mathrm{f}}=0,1,2$; the number of shown realizations if $4000$. Orange lines indicate linear fit of the distributions, which slopes give the normalised pressure $P_{\mathrm{tot}} = 0.38006 - 0.18684 \, \Delta \ell$ shown in (b) for $n=25$, where black circles are for the pressure computed for the data points without forks and empty squares with forks for the values $\Delta \ell = 1.15-1.35$ for which forks appear; \edit{all reported values of $P_{\mathrm{tot}}$ are obtained by averaging over all values of $\Delta x_{k}$, i.e. as the orange lines in fig.~\ref{fig:denergy-darea-3}}. The number of realizations: $3500$ for $\Delta \ell=0.2$, $1000$ for $\Delta \ell=0.25-1.15$, and $>4000$ for $\Delta \ell=1.2-1.35$. Inset in (b) shows dependence of $P_{\mathrm{tot}}$ on the system size $n$. (c) \textit{Top}: pile volume $V$ vs $\Delta x_{k}$ for $n=25$, $\Delta \ell=1.3$ with inset for $c_{V}(\Delta \ell)$. \textit{Bottom}: temperature of roughness $T_{\mathrm{r}}$ vs $\Delta x_{k}$ for $\Delta \ell=1.3$, $n=25$, and number of realizations $4000$; inset shows $c_{T}(\Delta \ell)$. (d) \textit{Top}: fork $T_{\mathrm{f}}$ (solid line, solid circles) and partition $T_{\mathrm{f}}^{\mathrm{part}}$ (dashed line, squares) temperatures as functions of $\Delta \ell$ for $n=25$; we used $20$ volume bands for computing $T_{\mathrm{f}}$ and $20$ energy bands for $T_{\mathrm{f}}^{\mathrm{part}}$. \textit{Bottom}: probability density $\rho_{i}$ as a function of $U_{i}$ showing a transition from no-fork states (black circles) characterised by a second-order temperature $T_{\mathrm{r}}$ to states with forks (empty squares) characterised by a first-order temperature $T_{\mathrm{f}}$ for $n=25$, $\Delta \ell=1.3$, $\Delta x_{\mathrm{max}} = 0.3465$, number of energy bands $20$, and $4000$ realizations. \edit{For color coding in (a,c,d-\textit{top}) corresponding to different $\Delta x_{k}$ refer to the colorbar in fig.~\ref{fig:denergy-darea-3}, which is the same as in fig.~\ref{fig:energy-area}}.}
\end{figure*}

We start by computing \textit{pressure} $P_{\mathrm{tot}} = - \left(\partial U/\partial V\right)_{S_{\mathrm{f}}}$ from the slope in fig.~\ref{fig:denergy-darea-3} \edit{\footnote{The reason why $\delta V<0$ in fig.~\ref{fig:denergy-darea-3} is because it was defined in computations as $\delta V = V(\Delta x_k+\delta x)-V(\Delta x_k)$ for $\delta x>0$ and we know that, at least for $\Delta\ell <1.35$, the larger the amplitude of a random ball distribution the smaller volume as per fig.~\ref{fig:roughness}-top. Hence for $\delta x>0$ we have $\delta V <0$, while for $\delta x<0$ we have $\delta V > 0$.}}, i.e. when the fork entropy is fixed in \eqref{relation:thermodynamic-fundamental}: the resulting fig.~\ref{fig:pressure} demonstrates $P_{\mathrm{tot}}$ to be a linear function of $\Delta \ell$. In fig.~\ref{fig:denergy-darea-3} we provide only one slope for all $\Delta x_{k}$ because comparing those for different $\Delta x_{k}$ does not exhibit any variation despite the fluctuations from the mean being substantial, which is a common property of finite-size systems \cite{Callen:1985}. Hence pressure is independent of $S_{\mathrm{r}}$ thus a posteriori justifying the consideration of $P_{\mathrm{tot}}$ instead of $P$, while the latter along with $T_{\mathrm{r}}$ will be identified in due course. Also, in general, GM systems are characterized by slower convergence with the number of particles $\mathcal{N}$ compared to standard thermal systems \cite{Scott:1960}. However, inset in fig.~\ref{fig:pressure} demonstrates that we work in the regime when pressure becomes independent of the system size $n$ thus justifying the values of $n$ used in our computations \edit{\footnote{A nonlinear fit analysis of the data in this inset shows that within the variance pressure does approach a constant value as $P_{\mathrm{tot}} = \const_{1} + \const_{2}/n^{a}$, where the power $a > 1$ with the actual value depending upon how many points from the right are taken into account, e.g. for $\Delta \ell = 1.0$ and all points $n \ge 10$ we get $P_{\mathrm{tot}} = 0.173 + 1.646/n^{1.451}$ with variance $0.003$ (square root of the sum of squared residuals at all data points, normalized by the number of data points).}}. The pressure proves to be positive, which physically amounts to the fact that, in a gravity-driven cohesion, a GM elastic energy increase $\delta U > 0$ is accompanied with a decrease of the pile volume $\delta V < 0$ as we observe in fig.~\ref{fig:denergy-darea-3}. Even though the pile is not homogeneous (in particular, due to hydrostatics), if we consider it as an elementary GM volume, similar to a fluid element which is inhomogeneous due to internal fluctuations and external forces, we may ascribe single thermodynamic quantities to such a finite-size element \cite{Hill:1962,Hill:1963,Rowlinson:1987,Unruh:1994,Bustamante:2005,Ritort:2007,Tovbin:2019}. That is, instead of dealing with individual forces and displacements of each ball, an integral view of the energy variation is done via perturbing the entire pile volume thus leading us to the concept of pressure \edit{as a measure of average ball cohesion} $P_{\mathrm{tot}} = -\delta U/\delta V$ provided the variation is done under fixed fork entropy, cf. equation \eqref{relation:thermodynamic-fundamental}. Hence, even though the distribution of forces is nonuniform across the pile \footnote{Besides being inhomogeneous due to the physical setting in the gravity field, there is another source of inhomogeneity -- the slopes, which apparently act as non-reflecting (absorbing) BCs. The fact that a pile has an ``interface'' can be seen from a thought experiment of impossibility to embed a pile in an extended layer of balls since balls on the slopes of the pile are not subject to the same external forces as the balls in the interior. This thought experiment is analogous to embedding a drop of water in the bulk, so that its surface properties are lost. Our system is finite with all the pertinent to such systems effects, but we worked in the (thermodynamic) regime when size dependence of the key scaled variables, such as pressure, disappears (inset to fig.~\ref{fig:pressure}).}, $P_{\mathrm{tot}}$ is its integral characteristic because the conjugate variable -- volume $V$ -- is also an integral one. After all, thermodynamics is a macroscopic theory, i.e. requires smearing over microscopic inhomogeneous portions of matter. In other words, we are developing a statistical physics approach to justify a thermodynamical description. However, while being homogenized over a collection of microscopic elements, thermodynamics is still applicable to macroscopically inhomogeneous systems \cite{Hart:1959,Rowlinson:1993,Lavenda:1995,Gujrati:2012} such as ours due to being set in a gravity field. Lastly, fig.~\ref{fig:pressure} shows that the effect of forks on the pressure magnitudes is not significant as the fork-gas is dilute, but the variance of pressure becomes more substantial due to larger fluctuations of the force magnitudes around the forks \edit{\footnote{Physically, this happens because a ball in a fork in $j$-th layer sits on balls from layers $j-1$ and $j-2$ compared to a non-fork position, in which ball sits on balls from $j-1$ layer only; for this discreteness, we do not distinguish between forks of different angular amplitudes.}}, which affects energy fluctuations as well. The pressure variance also increases for small values of $\Delta \ell$ because the relative energy fluctuations could be substantial as allowed by the solution of the force problem, i.e. a close-to-hexagonal pile is mechanically stable for larger variations of forces due to indeterminacy of the friction coefficients, whereas the volume fluctuations decrease.

Next, although \textit{temperature of roughness} can be formally found by computing $T_{\mathrm{r}} = \left(\partial U/\partial S_{\mathrm{r}}\right)_{S_{\mathrm{f}},V}$ according to \eqref{relation:thermodynamic-fundamental}, the variations of $S_{\mathrm{r}}$ and $V$ turn out to be dependent due to the existence of an equation of state because in the limit $\varepsilon \rightarrow 0$ only the probability parameter $\Delta x_{k}$ controls the problem for a fixed $\Delta \ell$. Indeed, since $\delta S_{\mathrm{r}}$ is naturally related to $\delta x$ in view of the definition of $S_{\mathrm{r}}$, and $\delta x$ inevitably generates variation of the pile volume $\delta V$ as well, which in turn is responsible for the origin of pressure $P_{\mathrm{tot}}$, we conclude that $T_{\mathrm{r}}$ and $P_{\mathrm{tot}}$ must be connected via an equation of state, that can be gleaned from plots in fig.~\ref{fig:roughness}. Namely, since the pile volume behaves \textit{as} $1-V \approx c_{V} \Delta x_{k}^{2}$ and the roughness temperature \textit{as} $T_{\mathrm{r}} \approx c_{T} \Delta x_{k}^{2}$, by variation of volume $V$ and entropy $S_{\mathrm{r}}$ with $\delta x$,
\begin{align}
\delta V = - 2 c_{V} \Delta x_{k} \delta x, \quad \delta S_{\mathrm{r}} = \frac{n-2}{\Delta x_{k}} \delta x,
\end{align}
from the fundamental thermodynamic relation \eqref{relation:thermodynamic-fundamental} for constant $S_{\mathrm{f}}$ we get
\begin{align}
\delta U = - \left[\frac{n-2}{2 (1-V)} T_{\mathrm{r}} + P\right] \delta V \equiv - P_{\mathrm{tot}} \delta V;
\end{align}
here the expression in square brackets is independent of $\Delta x_{k}$ and reduces to the equation of state $P_{\mathrm{tot}} = P + \frac{n-2}{2} \frac{c_{T}}{c_{V}}$, \edit{most accurate for a pile without forks since with the appearance of the latter the pressure variance increases as per fig.~\ref{fig:pressure}}. \edit{The factors $c_{V}$ and $c_{T}$ are found via fitting to the data and are functions of $\Delta \ell$, which may appear as a dependence on a microscopic parameter, but together with $n$ it defines a macroscopic footprint of the pile; hence, it is more convenient to think of the dependence of macroscopic parameters on $n$ only as elaborated below.} Notably, the behavior $1-V \approx c_{V} \Delta x_{k}^{2}$ demonstrates that increasing entropy $S_{\mathrm{r}}$, cf. plot $V(\Delta x_{k})$ in fig.~\ref{fig:roughness}-top, and appearance of forks, cf. fig.~\ref{fig:extensivity}, both contribute to compactification of the pile \footnote{However, the energy deviates in both directions away from $U_{\mathrm{reg}}$, which is due to balls being squeezed more on average.} somewhat similar to GM shaking experiments \cite{Scott:1960,Paillusson:2012}.
\edit{With these considerations, roughness temperature is computed from $T_{\mathrm{r}} = \left(\delta U/\delta S_{\mathrm{r}}\right)_{S_{\mathrm{f}}}$, where as before we compute pile realizations for $\Delta x_k$ and $\Delta x_k + \delta x$ with a sufficiently small $\delta x$, so that the pile geometric configurations are very close with the same number of forks and hence fork entropy $S_{f}$.} Also, the behavior $T_{\mathrm{r}} \approx c_{T} \Delta x_{k}^{2}$ proportional to the square of average deviations $\Delta x_{k}^{2}$ is analogous to that in the kinetic theory of gases where temperature is proportional to the average kinetic energy of molecules. Since (dimensional) energy of a regular pile scales as \edit{$\overline{U}_{\mathrm{reg}} \sim n^{3.009 \pm 0.004}$} and the entropy of roughness $S_{\mathrm{r}} \sim n$, the dimensional temperature of roughness scales as $\overline{T}_{\mathrm{r}} \sim n^{2}$ in the thermodynamic limit of large $n$. Correspondingly, because volume of a regular pile scales as $\overline{V}_{\mathrm{reg}} \sim n^{2}$, and the normalized pressure $P_{\mathrm{tot}}$ is constant for sufficiently large $n$, cf. inset in fig.~\ref{fig:pressure}, the dimensional pressure behaves as a hydrostatic pressure $\overline{P}_{\mathrm{tot}} \sim n$ in the thermodynamic limit thus demonstrating the expected macroscopic inhomogeneity of GM due to gravity and justifies the use of the term ``hydrostatic'' throughout this work.

Finally, with the appearance of forks, we may compute the corresponding \textit{fork temperature} $T_{\mathrm{f}} = \left(\partial U/\partial S_{\mathrm{f}}\right)_{S_{\mathrm{r}},V}$. It can be found from equation \eqref{relation:thermodynamic-fundamental} in two ways: either (i) by sorting out the data in volume bands, so that volume is kept constant $\d V \approx 0$, or (ii) by comparing energies with different fork entropies and volumes since $P_{\mathrm{tot}}$ is known; both approaches give the same result. The data indicate that, within the variance, for every $\Delta x_{\mathrm{max}}$ the temperature $T_{\mathrm{r}}$ does not show any trend with the fork entropy $S_{\mathrm{f}}$. Therefore, we take for $T_{\mathrm{f}}$ the average over all fork entropies $S_{\mathrm{f}}$, which proves to be negative. Since our system's elastic energy is at the minimum of the energy functional, the negative fork temperature has nothing to do with negative temperatures observed in common systems in non-equilibrium states \cite{Ramsey:1956,Landau:1969,Callen:1985,Leff:2021}. Negative temperature also does not contradict the second law of thermodynamics $\d S_{\mathrm{f}} \ge \d Q/T_{\mathrm{f}}$, i.e. temperature is allowed to be negative, and in our case simply indicates how much energy $\d U = \d Q$ is absorbed from the energy reservoir when the fork-gas ``condensates'' into the pure glassy state with decreasing $\Delta \ell$ and/or $\Delta x_{k}$, cf. fig.~\ref{fig:energy-area}. It seems that with appearance of forks our system behaves as a mixture of a dilute fork-gas and a disordered glass-like lattice. Despite the fork-gas being dilute, $T_{\mathrm{f}}$ is order of magnitude higher than $T_{\mathrm{r}}$, cf. figs.~\ref{fig:roughness} and \ref{fig:temperature}, which can be understood based on the following considerations.

\begin{figure*}[ht!]
	\setlength{\labelsep}{0.5mm}
	\centering
	\sidesubfloat[]{\epsfig{figure=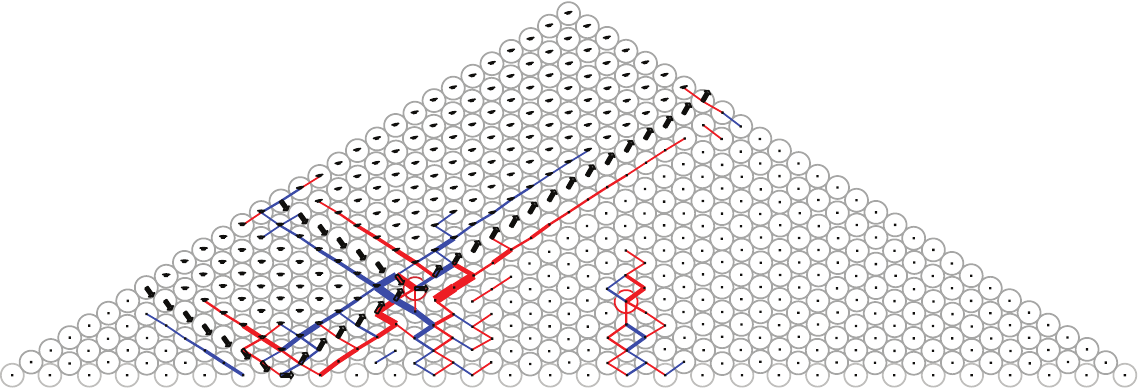,height=2.25in}\label{fig:rheology}} \setlength{\labelsep}{-4.85mm} \\[0.5cm]
    \sidesubfloat[]{\epsfig{figure=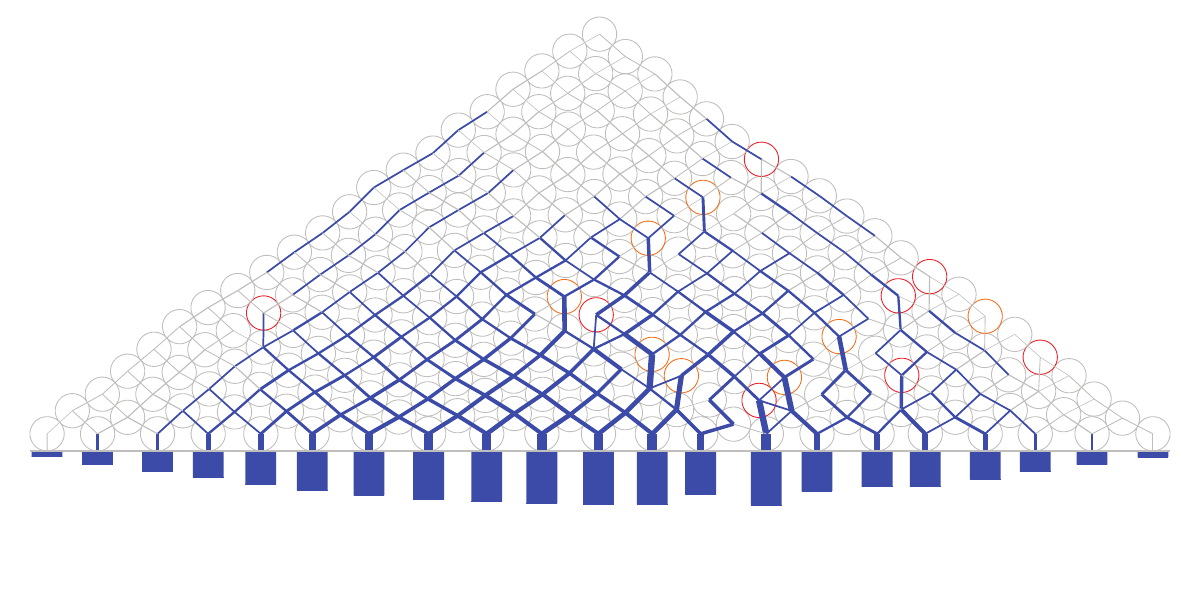,height=1.75in}\label{fig:arc-normal}}
	\sidesubfloat[]{\epsfig{figure=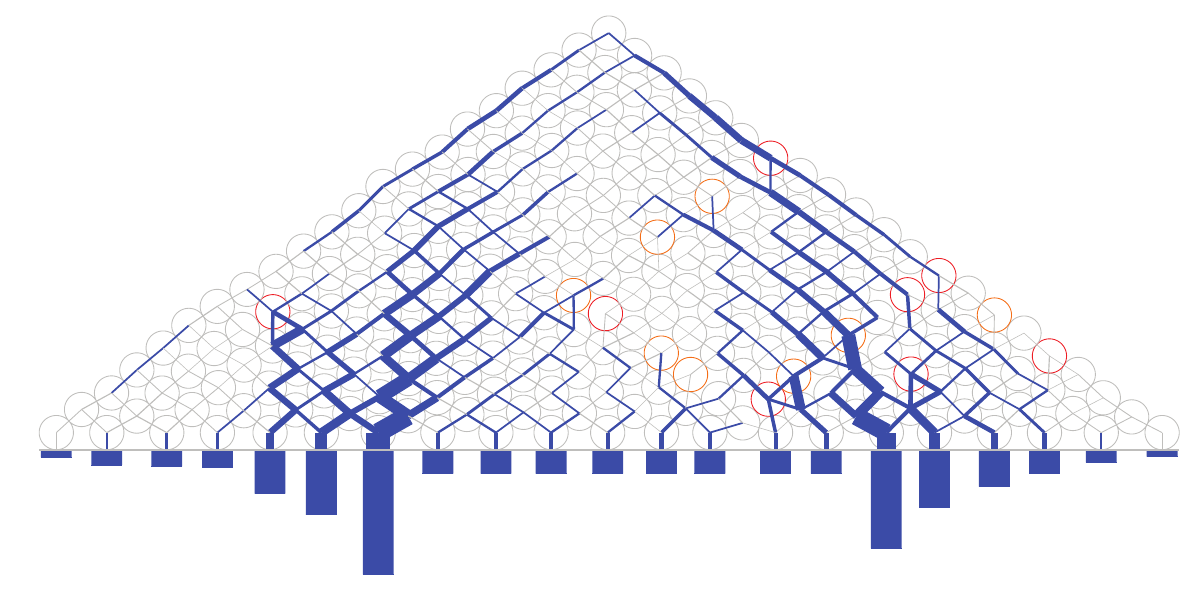,height=1.75in}\label{fig:arc-less}}
\caption{\setlength{\rightskip}{0pt} \setlength{\leftskip}{0pt} Rheology: (a) perturbations of ball positions and force amplitudes of an irregular pile ($n=30$, $\Delta \ell=1.35$) are achieved by shifting the eighth (from the left) ball in the first layer by $5\%$ of $\Delta x_{\mathrm{max}}$; all displacements are marked with black segments (magnified for visibility); the force perturbations are normalized by the amplitude of fluctuations with blue denoting positive variations and red negative ones; (b) distribution of stresses in a pile under normal conditions -- shown with the bars is the distribution of normal forces on the substrate; stress strength is reflected in the thickness of the blue lines connecting the balls; forks are marked red. In (c) the pile is perturbed with the normal forces acting on the substrate in the middle being $70\%$ of the average amplitude in (b), while the total pile weight is kept constant resulting in larger amplitudes to the left and right from the center qualitatively analogous to the procedure in \cite{Nadukuru:2012}. The conditions for (b,c): $n=21$, $\Delta \ell =1.2$, $\Delta x_{\mathrm{max}}=0.4$.}
\end{figure*}
From the point of view of the statistical physics, i.e. if we plot the Boltzmann-Maxwell probability density $\rho_{i}$ as a function of energy $U_{i}$ (bottom of fig.~\ref{fig:temperature}), we observe a transition from the \textit{no-fork} to \textit{fork} state residing in a relatively narrow range of energies near that of a regular pile, which is \edit{well} described by a parabolic function $\ln{\rho_{i}} = A-B \Delta U_{i}^2 / \Delta x_{k}^{2}$, where $\Delta U_{i} = U_{i}-U_{\max}$ \edit{is the energy departure from maximum energy $U_{\max}$ for configurations without forks (cf. parabola in fig.~\ref{fig:temperature}-bottom)}, $A$ is a normalization constant, and the curvature $B$ depends on $\Delta \ell$, but not on $\Delta x_{k}$. \edit{While the parabolic dependence of $\ln{\rho_{i}}$ on $\Delta U_{i}$ and scaling with $\Delta x_{k}^{-2}$ were originally motivated by the numerical data in fig.~\ref{fig:temperature}-bottom, they found further physical justification from the following considerations.} Because of the parabolic behavior, the roughness temperature $T_{\mathrm{r}}$ is of \textit{second-order} \cite{Ingarden:1994}, the origin of which can be explained in the classical thermodynamics terms. Namely, using the conservation of energy $U$ for the pile and gravitational energy $E_{g}$ of the reservoir, $E = U + E_{g}$, the Boltzmann probability $p_{i}$ of occurrence of the situation when a pile is in the microstate $i$ with energy $U_{i}$ in the ensemble is proportional to the number of states $\Omega_{g}(E-U_{i})$ accessible to the reservoir under these conditions normalized by the total number of states $\Omega_{t}(E)$ of the total system, i.e. $p_{i} = \Omega_{g}(E-U_{i})/\Omega_{t}(E)$, which is a \textit{seminal relation} in canonical formalism \cite{Callen:1985} and does not rely upon additivity of entropy. Thus, Taylor expanding to second order:
\begin{align}
\label{temperature:roughness-second}
\!\!\!\!\ln{p_{i}} \approx \left.\ln{\frac{\Omega_{g}}{\Omega_{t}}}\right|_{E} - \left.\frac{\partial \ln{\Omega_{g}}}{\partial E_{g}}\right|_{E} \!\!\!\!\cdot U_{i} + \frac{1}{2} \left.\frac{\partial^{2} \ln{\Omega_{g}}}{\partial E_{g}^{2}}\right|_{E} \!\!\!\!\cdot U_{i}^{2},
\end{align}
where $\left.\frac{\partial \ln{\Omega_{g}}}{\partial E_{g}}\right|_{E} = 1/T_{\mathrm{r}}^{(1)}$ would produce the usual first-order temperature of roughness $T_{\mathrm{r}}^{(1)}$ \edit{\footnote{The situation is analogous to the Earth atmosphere, which obviously exchanges energy with the Earth gravitational field. While constant in its composition and suppose, for simplicity, isothermal as in Newton's derivation (leading to an exponential form for pressure distribution with the altitude), the atmosphere may fluctuate in its microstate energy, volume, entropy, etc. as in the canonical description. For atmosphere, one can also perform the standard canonical analysis along the lines of equation \eqref{temperature:roughness-second} in the manuscript, for which (as we see) one does not need to know the expression for $\Omega_{g}$ and calculate the corresponding derivative $\left.\partial \ln(\Omega_g)/\partial E_g\right|_{E}$, whereas the latter is $1/T_r^{(1)}$. Hence, the employed canonical description for GM is valid with the key difference from standard thermodynamics is that the third law does not apply to this total system as the reservoir and the GM pile have different nature.}}, which is \textit{infinite} in the considered here case since the number of accessible states for each ball in the first layer diverges $\Delta x_{\mathrm{max}}/\varepsilon \rightarrow \infty$, whereas $\left.\frac{\partial^{2} \ln{\Omega_{g}}}{\partial E_{g}^{2}}\right|_{E} = 1/T_{\mathrm{r}}$ gives our finite second-order temperature of roughness $T_{\mathrm{r}}$ and thus explains why it is order-of-magnitude lower compared to the fork temperature $T_{\mathrm{f}}$. Comparing the expression for $\ln{p_{i}}$ resulting from theory \eqref{temperature:roughness-second} and $\ln{\rho_{i}}$ from experiments, keeping in mind that the multiplicity factor relating $p_{i}$ and $\rho_{i}$ as well as the shift of the base energy $E \rightarrow E -U_{\mathrm{max}}$ in \eqref{temperature:roughness-second} do not affect the definition of temperature, we conclude that the factor in $T_{\mathrm{r}} \approx c_{T} \Delta x_{k}^{2}$ is $c_{T} = -\frac{1}{2 B}$, cf. inset in fig.~\ref{fig:roughness}-\textit{bottom}.

With the appearance of forks, the behavior of $\ln{\rho_{i}}$ in fig.~\ref{fig:temperature}-\textit{bottom} changes from parabolic to linear $\ln{\rho_{i}} = \beta_{0} - \beta U_{i}$ and operates on a much wider range of energies as expected from fig.~\ref{fig:energy-area}. Therefore, it is the latter regime where we recognize the thermodynamic behavior common in thermal systems \cite{Reif:1965,Callen:1985}, i.e. when $\beta_{0}$ is related to the partition function $Z = \sum_{i}{e^{-\beta U_{i}}}$ as $\beta_{0} = - \ln{Z}$ and $\beta$ is the usual \textit{thermodynamic beta} giving us the ``partition'' version of the fork temperature $T_{\mathrm{f}}^{\mathrm{part}} = \beta^{-1}$, cf. fig.~\ref{fig:temperature}-\textit{top}. While the order of magnitude of the fork temperature $T_{\mathrm{f}}$ computed with statistical methods and $T_{\mathrm{f}}^{\mathrm{part}}$ computed with the Boltzmann partition function are the same, there are some differences, which are due to non-ergodicity of the system and hence departure from a Boltzmann factor $e^{-\beta U_{i}}$ embedded in the corresponding partition function description hinged on the assumption of ergodicity \cite{Reif:1965,Callen:1985}. Indeed, our system does not necessarily satisfy the fundamental statistical postulate that a system attends every part of the phase space with equal probability also known as the ergodicity hypothesis: in our case, local minima of the $U$-functional may be located at the same energy level, but well-separated in the phase-space landscape. Non-ergodicity is not surprising since almost none of the classical models of statistical mechanics satisfy the hypothesis -- nevertheless, this fact does not invalidate statistical mechanics description of the system as the ergodic hypothesis is not required when making statements about an ensemble, not a particular system \cite{Landau:1969}. Moreover, the ergodic hypothesis is usually needed to make statements about time-averages, which we do not have to deal with as our system is static.

\vspace{-0.35cm}

\section{Rheology} \label{sec:rheol}

\vspace{-0.35cm}

Perturbation of a regular (isosceles triangular) pile leads to a set of hyperbolic equations for the ball position deviations $\boldsymbol{\xi} = (\xi,\eta)$ from a regular isosceles pile with angle $\phi$:
\begin{align}
\label{eqs:hyperbolic-regular}
\boldsymbol{\xi}_{xx} - \cot^{2}{\!\phi} \,\, \boldsymbol{\xi}_{yy} = 0,
\end{align}
cf. Materials and Methods. Since balls displacements are translated hyperbolically along the characteristics $y \pm \cot{\!\phi} \, x$ of the wave equation \eqref{eqs:hyperbolic-regular} from the first layer on, cf. fig~\ref{fig:rheology}, the displacement fluctuations at every layer are determined by $\Delta x_{k}$ at the first layer and they are homogeneously distributed across the entire pile. Forks, in fact, originate from the intersection of two characteristics carrying sufficiently large perturbations from the first layer. Based on this consideration, it is straightforward to show that forks appear above transition boundary in the $(\Delta \ell, \Delta x_{k})$-plane defined by $\Delta x_{k} \ge (\sqrt{3}-1) - \frac{1}{2} \Delta \ell$ in the limit $n \rightarrow \infty$. Since forks result from a collective behavior of balls, they can be regarded as \textit{quasiparticles} per the term coined by Landau \cite{Landau:1933}. In view of that and since forks are almost non-interacting \edit{being diluted in the pile}, they behave as an ideal gas, in particular analogous to phonons which are also quasiparticles \cite{Iben:2012} \edit{or even dilute granular gas \cite{Luding:2001,Weele:2008}}. Because the fork-gas is dilute, its effect on the homogeneity of the ball position fluctuations is small, which is also reflected on the behavior of pressure in fig.~\ref{fig:pressure}, but $T_{\mathrm{f}}$ is higher than $T_{\mathrm{r}}$ due to larger relative energy variances in the presence of forks, as seen in fig.~\ref{fig:energy-area}.

With the appearance of forks, wave-like propagation of initial conditions at the first layer becomes analogous to wave propagation in random media with scattering on impurities. Indeed, the hyperbolic character of the geometric part of the problem can be gleaned in fig.~\ref{fig:rheology}, in which we can observe how displacing eighth ball in the first layer leads to propagation of the displacement perturbation along the characteristics defined by $\cot{\phi}$. Moreover, once a characteristic hits a fork, it scatters on it emanating another characteristic to the left.

Similarly, perturbation of the regular pile yields a rheological equation relating stress $\boldsymbol{\sigma} = \nabla \left(\delta \boldsymbol{N}\right)$ and strain $\nabla \boldsymbol{\xi}$ tensors (cf. Materials and Methods), given here in a particular form:
\begin{align}
\label{eqn:rheology}
\begin{pmatrix}
  A_{1} & A_{2} \\
  B_{1} & -B_{2}
\end{pmatrix}
\begin{pmatrix}
  \grave{\sigma} \\
  \acute{\sigma}
\end{pmatrix} = - W
\begin{pmatrix}
\xi_{x} \left[1-\frac{1}{A_{1}\cos{\phi}}\right] \\
  \xi_{y}
\end{pmatrix}
\end{align}
where the unknowns -- $\grave{\sigma} = - \delta \grave{N}_x \sin{\phi} - \delta \grave{N}_y \cos{\phi}$ and $\acute{\sigma} = \delta \acute{N}_x \sin{\phi}-\delta \acute{N}_y \cos{\phi}$ -- are the scalar projections of $\boldsymbol{\sigma}$ at the left and right contact points correspondingly, whereas $A_{i}=\cos\phi+\mu_{i}\sin\phi$ and $B_{i}=\sin\phi-\mu_{i}\cos\phi$ are the matrix elements. Since the determinant $\det=(1-\mu_1\mu_2)\sin\phi\cos\phi+(\mu_1+\mu_2)(\sin^2\phi-\cos^2\phi)$ of the left-hand side of system \eqref{eqn:rheology} may change the sign depending on the values of frictions coefficients $\mu_{1,2}$ and the local angle $\phi$, the characteristic type of the rheology changes between hyperbolic and elliptic; notably, hyperbolicity at macroscopic level was concluded previously based on different considerations \cite{Blumenfeld:2004}. While fig.~\ref{fig:rheology} also illustrates how force perturbations may propagate hyperbolically along the characteristics, they also exhibit diffusive character, which happens only around certain nodes: as suggested by equation \eqref{eqn:rheology} such selectivity is dictated by the actual local friction coefficients and, in general, by the background fields of hydrostatics $\boldsymbol{N}(\boldsymbol{x})$ and angles $\phi(\boldsymbol{x})$ of an irregular pile. The interplay of hyperbolic and elliptic characteristics is most apparent in the origin of force chains as in fig.~\ref{fig:arc-less}, which shows a perturbation of a pile formed under normal conditions in fig.~\ref{fig:arc-normal}, i.e. when the elastic energy functional is minimized and the pressure distribution under the pile is a bell-like as expected \cite{Vanel:1999,Nadukuru:2012}. Near the forks, the stresses can considerably deviate from those at the neighboring balls due to relaxation as reflected in fig.~\ref{fig:energy-area}. Both force chains and arching are manifestations of the long-range non-diffusive propagation of perturbations, which is obviously mediated with the hyperbolic character of the underlying rheological equations \eqref{eqn:rheology}. However, we also observe eventual decay of force chains as in fig.~\ref{fig:arc-less}, which is due to diffusive character of \eqref{eqn:rheology} occurring at some nodes.

With the above understanding of characteristic types of the geometric and force problems, we are in a position to interpret fig.~\ref{fig:extensivity}, which reveals some important physics. Namely, for \textit{small volumes} $V$ the behavior of the fork entropy $S_{\mathrm{f}}$ is close to extensive, which is because the probability of large fluctuations in the first layer is proportional to $n$ and hence the immediate number of forks produced by the intersection of characteristics emanating from the first layer is $\propto n^{2}$ because dissipation of hyperbolic characteristics (and thus force chains evident from figs.~\ref{fig:arc-normal} and \ref{fig:arc-less}) is weak since there are less chances of diffusive spots. For \textit{large volumes} $V$, the process of dissipation, mediated by switching from hyperbolic to elliptic behavior at the level of force description, takes over and therefore the slope in fig.~\ref{fig:extensivity} decreases and the entropy $S_{\mathrm{f}}$ becomes substantially non-extensive.

\vspace{-0.45cm}


\section{Discussion}

\vspace{-0.25cm}

Thermodynamics is a science of the relationship between temperature, available forms of energy, and properties of matter. In the present work, we establish such an interrelation for GM in a glassy state formed in a natural setting with cohesion mediated by gravity. Using decoupling of geometric and force problems, we demonstrated that the geometric fabric $\boldsymbol{\xi}$ of GM has a hyperbolic character with two sources of uncertainty and hence entropy: the \textit{roughness} of the substrate, on which the first layer balls are positioned, and \textit{forks}, appearing when sufficiently large perturbations at the first layer propagate along two intersecting characteristics. The phase transition from no-fork to fork states is marked by larger variations of the pile energy rather than volume: therefore, despite being dilute, the fork-gas is characterized by significant temperatures. While geometrically homogeneous, at the force level  the GM is inherently inhomogeneous macroscopically for being set in a gravity field and microscopically due to the emergence of force chains. The latter proved to originate from the mixed hyperbolic-elliptic type of GM rheology. Smearing out the microscopic inhomogeneities -- individual ball displacements and forces acting upon them -- we showed that classical thermodynamic description is adequate for an ensemble of pile realizations, i.e. the averaged macroscopic properties ($T_{\mathrm{r}},T_{\mathrm{f}},P_{\mathrm{tot}},V$) are insensitive to the microscopic details of particular pile realizations, though with a number of important caveats differentiating GM from familiar systems. Namely, though without forks GM has a well-defined equation of state, it involves second-order temperature only. First-order temperature appears with the birth of forks forming a dilute and hence ideal gas of weakly interacting quasiparticles. Moreover, the fork entropy turns out to be non-extensive and macroscopic properties such as $T_{\mathrm{r}}$, $T_{\mathrm{f}}$, $P_{\mathrm{tot}}$, normally intensive in standard thermodynamic systems, scale with the system size. In the case of pressure, this is not entirely surprising as $P_{\mathrm{tot}} \sim n$ similar to hydrostatics and is a reflection of macroscopic inhomogeneity of the system.

With regard to being testable, the fact that our geometric and force problems are decoupled makes connection to real-world experiments, in particular the ones for 2D balls (disks) \cite{Majmudar:2005,Geng:2001,Bililign:2019}, straightforward. Namely, given the number of balls $n$ in the first layer, measured values of $\Delta \ell$ and $\Delta x_{k}$, we may immediately tell what is the temperature of roughness $T_{\mathrm{r}}$, total pressure $P_{\mathrm{tot}}$ (and hence $P$), as well as predict the volume $V$ of the pile and the number of forks along with the associated fork temperature $T_{\mathrm{f}}$. Of course, all these predictions concern the averaged properties for an ensemble of piles, whereas a particular pile realization falls within the corresponding variances. Accordingly, a ``thermometer'', analogous to the one we use to measure temperature of thermal systems, would be a device identifying the properties $n$, $\Delta \ell$, $\Delta x_{k}$ of the first layer. The amazing feature of GM is that these properties in the first layer dictate the geometry of the entire pile, including the appearance of forks. The established interrelation of all the introduced thermodynamic quantities not only enables us to predict the values of some thermodynamic variables based on the knowledge of the others in analogy to an ideal gas, for example, but most importantly reveals the internal purport of GM, which proves to be very different from common thermal systems. In particular, it urges us to change the standard paradigm of how we think about thermodynamics of complex systems in (unjustified) analogy to thermal systems: since GM properties are not contagious -- due to the absence of the zeroth law of thermodynamics for GM systems a second pile brought in contact with the first one is not going to acquire its thermodynamic properties as the thermometer put in contact with a macroscopic body -- a thermometer-like device must be a bit more sophisticated to glean the pile properties as we hinted above. This shifted paradigm may apply to many other systems more complex compared to familiar condensed matter.

The reader may ask what happens in general GM systems, which are 3D and involve grains of different shapes and sizes. What we constructed here is the equivalent to the Ising model of ferromagnetism, which exhibits the properties common of realistic GM such as force chains and arcs, thus providing a firm footing for understanding GM as a state of matter. The developed theoretical framework demonstrates that GM does not behave as fluids or solids, but is rather a rich collection of various phenomena. Adding further details such as varying shapes and sizes will enlarge the parameter space to work with thus \edit{potentially enlarging/modifying the relevant thermodynamic variables and} offering a path for future explorations. And while we were not quite able ``to see a world in a grain of sand'' (W. Blake), with this model we uncovered the fascinating story of intertwined thermodynamics and rheology of glassy GM thus setting a reference point for understanding GM in general and providing a basis for greater enjoyment when walking on sandy beaches.

\appendix*

\vspace{-0.25cm}

\section{MATERIALS AND METHODS}

\vspace{-0.25cm}

\subsection{3-ball problem} \label{subsec:3-balls}

\begin{figure}[!hbt]
    \centering
      \includegraphics[width=0.75 \textwidth]{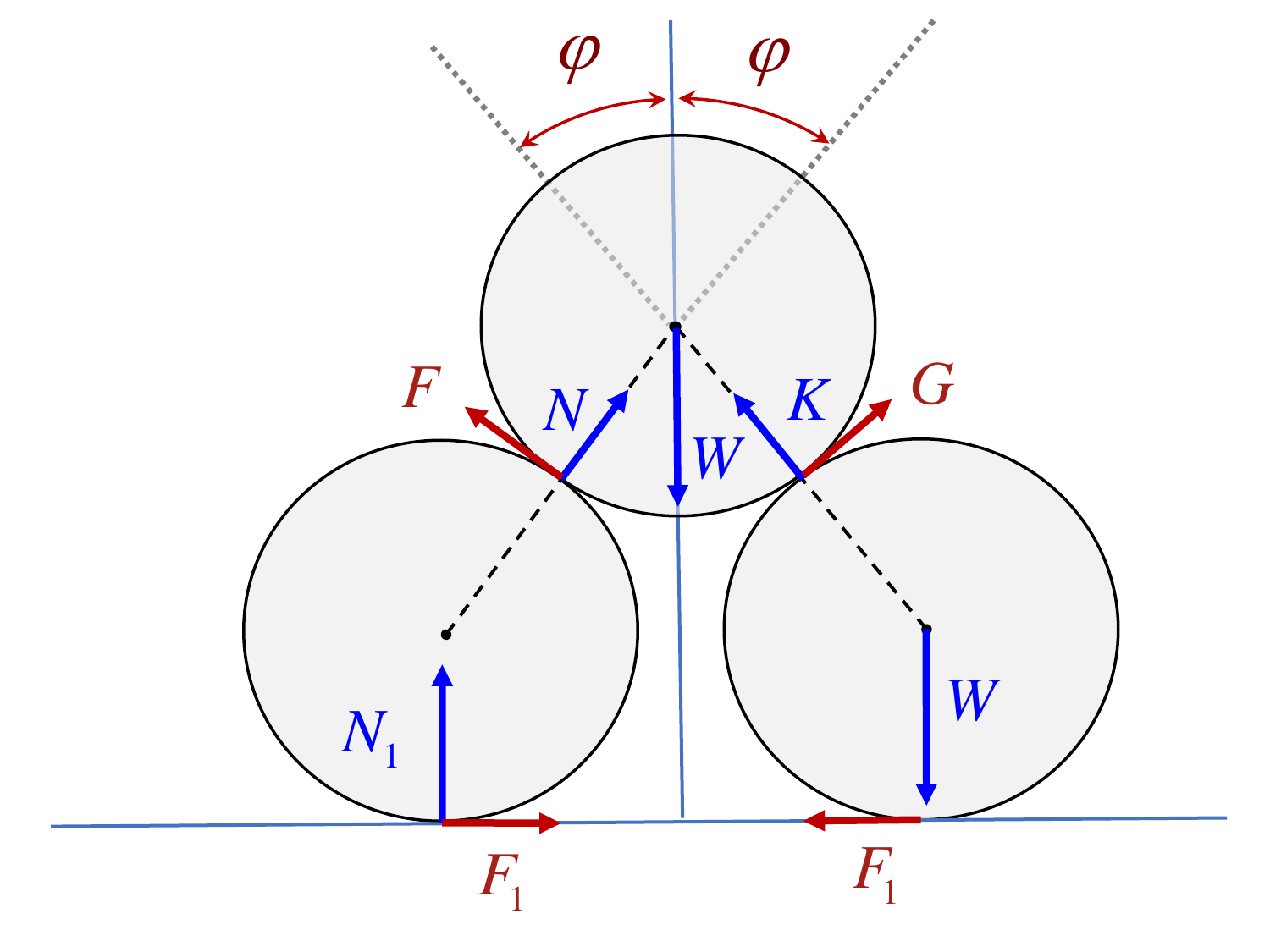}
    \caption{\setlength{\leftskip}{0pt} \setlength{\rightskip}{0pt} The problem of stability of three balls. The static friction coefficient between balls and substrate is assumed to be the same $\mu_{s}$. \hspace{3cm}\hspace{4cm}}
    \label{fig:setup:three-balls}
\end{figure}
The first observation we can make about the pile of three balls is that in the symmetric case the upper ball can only slide, whereas the lower balls may either slide or roll. It is the friction between the upper and lower balls that holds a pile together -- without it the upper ball would slide down under the gravity thus making the lower ones to roll.

Let us start with analyzing the situation when the stability is lost through sliding. If the balls are lying on a flat surface then from balancing the vertical components of forces acting on the upper ball, cf. fig.~\ref{fig:setup:three-balls}, we find
\begin{align}\label{NW}
y: \ \frac{W}{2} - N \cos\phi - F\sin\phi = 0,
\end{align}
where $N,F,W$ denote the absolute values of the vector forces. From the balance of torques acting on the lower left ball, which guarantees no ball rolling, we get
\begin{align}\label{FF1}
F_1=F,
\end{align}
which is equivalent to
\begin{align}
\mu N=\mu_1 N_1,
\end{align}
where $\mu=\mu(\phi)$ and  $\mu_1=\mu_1(\phi)$. From the balance of forces acting on that ball we get two equations for the corresponding projections
\begin{subequations}
\label{NF}
\begin{align}
x&: & N\sin\phi &= F\cos\phi+F_1=F(1+\cos\phi),\\
y&: & N_1 &= W+N\cos\phi+F\sin\phi,
\end{align}
\end{subequations}
where we used \eqref{FF1} and took into account that the normal and friction forces acting on the lower ball are equal to those acting on the upper ball, cf. fig.~\ref{fig:setup:three-balls}, but have opposite signs. From \eqref{NW} and \eqref{NF} we obtain
\begin{subequations}
\begin{align}
 &F=\frac{W}{2}\frac{\sin\phi}{1+\cos\phi},\\
 &N=\frac{W}{2},\\
 &N_1=\frac{3W}{2},
\end{align}
\end{subequations}
and hence the ratios of friction and normal forces are
\begin{subequations}
\begin{align}
&\mu=\frac{\sin\phi}{1+\cos\phi}, \\
&\mu_1=\frac{\sin\phi}{3(1+\cos\phi)}=\frac{1}{3}\mu.
\end{align}
\end{subequations}
Therefore, if the static friction coefficient $\mu_{s}$ (for simplicity assumed to be the same between balls and between balls and substrate) is such that $\mu<\mu_{s}$, then there is no sliding at both contacts and the pile is stable. If, on other hand, $\mu>\mu_{s}$, but $\mu_{1} < \mu_{s}$, i.e. $\mu > \mu_{s} > \frac{1}{3} \mu$, the upper ball is sliding, whereas the lower balls are rolling. Finally, if $\mu > \frac{1}{3} \mu > \mu_{s}$, then the pile stability is lost by sliding at all contact points.

\subsection{Quadratic programming problem} \label{subsec:linear-programming}

\vspace{-0.25cm}

\begin{figure}[h!]
	\setlength{\labelsep}{-3.1mm}
	\centering
    \epsfig{figure=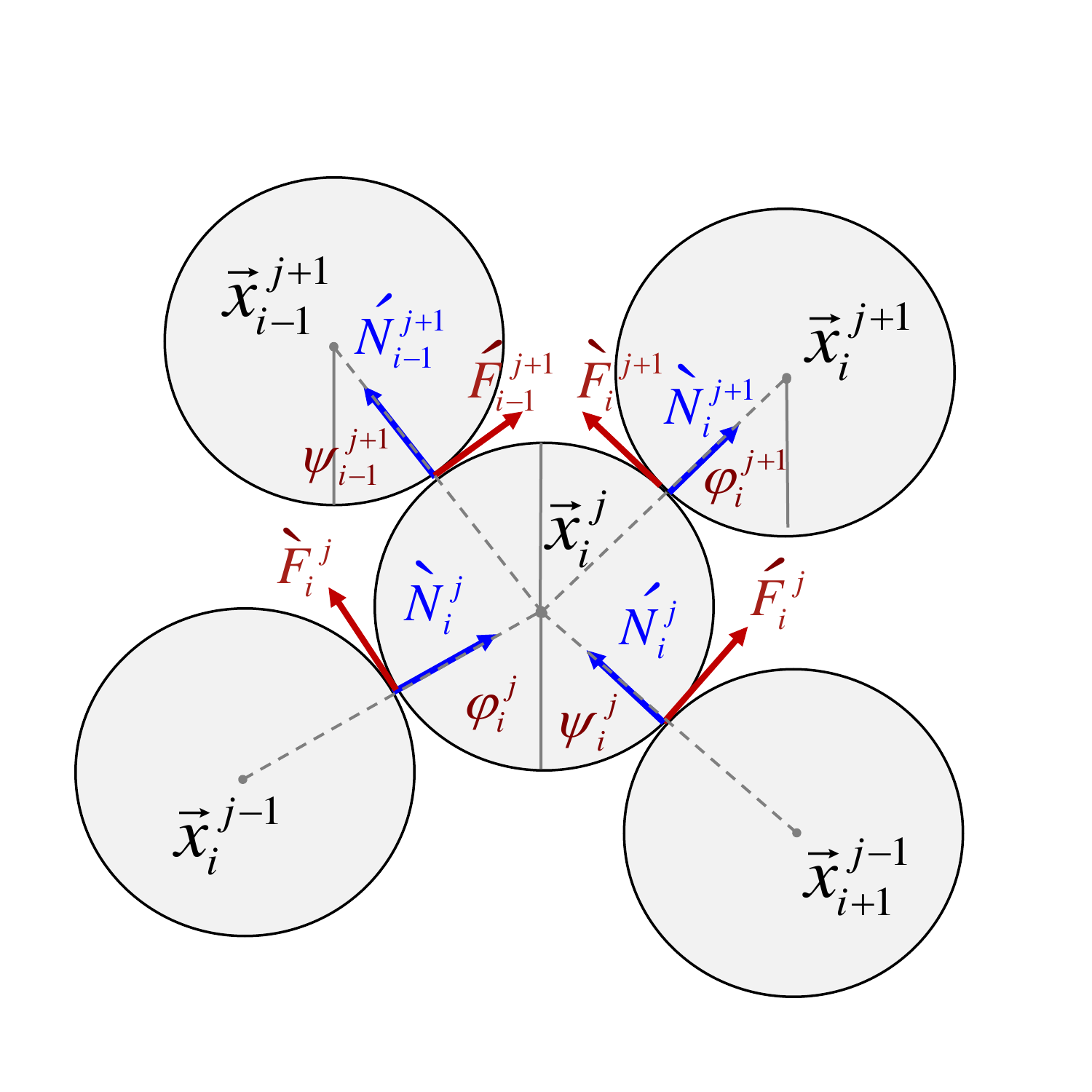,height=2.75in}
\caption{Notations for forces.}\label{fig:force-notations}
\end{figure}

The $i$-th ball in the $j$-th layer is correspondingly indexed with $(i,j)$, where $j=1$ for the bottom and $j=n$ for the top layer, while $i$ varies from $1$ (on the left) to $n-j+1$ in the $j$-th layer. Then, the forces are denoted as follows, cf. fig.~\ref{fig:force-notations}: $\grave{N}_i^j$ ($\grave{F}_i^j$) is the normal (friction) force at a contact point between balls $(i,j)$ and $(i,j-1)$, or $(i+1,j-2)$ in the case of a fork; $\acute{N}_i^j$ ($\acute{F}_i^j$) is the normal (friction) force at a contact point between balls $(i,j)$ and $(i+1,j-1)$, or $(i+1,j-2)$ in the case of a fork. For every ball located at $(x_{i}^{j}, y_{i}^{j})$, $\phi_{i}^{j}$ is the ``left'' angle between vertical and the direction to the neighbouring ball on the level $(j-1)$ to the left and $\psi_{i}^{j}$ is the ``right" angle between vertical and the direction to the neighbouring ball on the level $(j-1)$ to the right, cf. fig.~\ref{fig:force-notations}.

The governing equations include a projection of forces on the horizontal and vertical axes as well as a balance of the torques, e.g. for the layer $1<j<n$ without forks
\begin{widetext}
\begin{subequations}
\begin{equation}
\label{eqs:i1j}
i=1: \begin{cases}
&\grave{N}_1^j \sin\phi_1^j-\grave{F}_1^j\cos\phi_1^j-\acute{N}_1^j\sin\psi_1^j+\acute{F}_1^j\cos\psi_1^j-\grave{N}_1^{j+1}\sin\phi_1^{j+1}+\grave{F}_1^{j+1}\cos\phi_1^{j+1}=0,\\
&\grave{N}_1^j \cos\phi_1^j+\grave{F}_1^j\sin\phi_1^j+\acute{N}_1^j\cos\psi_1^j+\acute{F}_1^j\sin\psi_1^j-\grave{N}_1^{j+1}\cos\phi_1^{j+1}-\grave{F}_1^{j+1}\sin\phi_1^{j+1}-W=0,\\
&\grave{F}_1^j-\acute{F}_1^j+\grave{F}_1^{j+1}=0,
\end{cases}
\end{equation}
\begin{equation}
\label{eqs:ij}
1<i<n-j+1: \begin{cases}
&\grave{N}_{i}^{j}\sin\phi_{i}^{j}-\grave{F}_{i}^{j}\cos\phi_{i}^{j}-\acute{N}_{i}^{j}\sin\psi_{i}^{j}+\acute{F}_{i}^{j}\cos\psi_{i}^{j} \\
&~~+\acute{N}_{i-1}^{j+1}\sin\psi_{i-1}^{j+1}-\acute{F}_{i-1}^{j+1}\cos\psi_{i-1}^{j+1}-\grave{N}_i^{j+1}\sin\phi_i^{j+1}+\grave{F}_i^{j+1}\cos\phi_i^{j+1}=0,\\
&\grave{N}_{i}^{j}\cos\phi_{i}^{j}+\grave{F}_{i}^{j}\sin\phi_{i}^{j}+\acute{N}_{i}^{j}\cos\psi_{i}^{j}+\acute{F}_{i}^{j}\sin\psi_{i}^{j} \\
&~~-\acute{N}_{i-1}^{j+1}\cos\psi_{i-1}^{j+1}-\acute{F}_{i-1}^{j+1}\sin\psi_{i-1}^{j+1}-\grave{N}_i^{j+1}\cos\phi_i^{j+1}-\grave{F}_i^{j+1}\sin\phi_i^{j+1}-W=0,\\
&\grave{F}_{i}^{j}-\acute{F}_{i}^{j}-\acute{F}_{i-1}^{j+1}+\grave{F}_i^{j+1}=0,
\end{cases}
\end{equation}
\begin{equation}
\label{eqs:i1j1}
i=n-j+1: \begin{cases}
&\grave{N}_{n-j+1}^j\sin\phi_{n-j+1}^j-\grave{F}_{n-j+1}^j\cos\phi_{n-j+1}^j-\acute{N}_{n-j+1}^j\sin\psi_{n-j+1}^j\\
&~~+\acute{F}_{n-j+1}^j\cos\psi_{n-j+1}^j
+\acute{N}_{n-j}^{j+1}\sin\psi_{n-j}^{j+1}-\acute{F}_{n-j}^{j+1}\cos\psi_{n-j}^{j+1}=0,\\
&\grave{N}_{n-j+1}^j\cos\phi_{n-j+1}^j+\grave{F}_{n-j+1}^j\sin\phi_{n-j+1}^j+\acute{N}_{n-j+1}^j\cos\psi_{n-j+1}^j\\
&~~+\acute{F}_{n-j+1}^j\sin\psi_{n-j+1}^j
-\acute{N}_{n-j}^{j+1}\cos\psi_{n-j}^{j+1}-\acute{F}_{n-j}^{j+1}\sin\psi_{n-j}^{j+1}-W=0,\\
&\grave{F}_{n-j+1}^j-\acute{F}_{n-j+1}^j-\acute{F}_{n-j}^{j+1}=0,
\end{cases}
\end{equation}
\end{subequations}
\end{widetext}
and similarly for the first and $n$-th layers. Also, for all layers the following inequality constraints
\begin{align}
\label{eqs:inequalities}
\grave{N}_{i}^{j}, \acute{N}_{i}^{j} &\ge 0, \quad j\in[2,n],  \quad i\in[1,n-j+1],
\end{align}
and similar ones for $j=1$ must be satisfied. In the case of a pile with forks, some dynamical equations are to be modified, because in the presence of a fork the ball in the $j$-th layer can touch the balls in $(j-2)$-th or $(j+2)$-th layers. Moreover the number of its neighbours may be 2,3,4,5. All these possibilities lead to the according modification of the system of dynamical equations. But for every given packing the set of governing equations is unambiguous.

The total system contains $\frac{n(n-3)}{2}$ undefined forces in the case of a regular packing and $\frac{n(n-3)}{2}-1$ for a pile with forks. Therefore, under the condition of minimization of the elastic energy, which under the assumption of small deformation of balls (large Young modulus) is proportional to the sum of the squares of all forces
\begin{align}
\label{energy:functional}
U=C \sum_{i,j} \Big[(\grave{N}_{i}^{j})^2+(\acute{N}_{i}^{j})^2+(\grave{F}_{i}^{j})^2+(\acute{F}_{i}^{j})^2\Big],
\end{align}
with $C$ being a factor defined by the Young modulus, the above problem becomes the one of quadratic programming \cite{Wright:2015}, i.e. minimization of the functional \eqref{energy:functional} quadratic in forces while varying all undefined forces and satisfying the constraints and inequalities \eqref{eqs:inequalities}. Because the energy $U$ in 2D is quadratic in forces, the minimum is unique, but may not exist for some variants of packing because inequalities \eqref{eqs:inequalities} may not be satisfied.

\subsection{Rheology}

Let us start with an ordered set of balls with coordinates $\overline{\bm{x}}_{i}^{j}=(\overline{x}_{i}^{j},\overline{y}_{i}^{j})$ as shown in figure~\ref{fig:force-notations}; obviously, the coordinates must satisfy
\begin{subequations}
\label{eqs:coordinate-constraints}
\begin{align}
\left(\overline{x}_{i}^{j} - \overline{x}_{i}^{j-1}\right)^{2} + \left(\overline{y}_{i}^{j} - \overline{y}_{i}^{j-1}\right)^{2} &= 4r^2, \\
\left(\overline{x}_{i}^{j} - \overline{x}_{i+1}^{j-1} \right)^{2} + \left(\overline{y}_{i}^{j} - \overline{y}_{i+1}^{j-1}\right)^{2} &= 4r^2,
\end{align}
\end{subequations}
where $r$ is the ball radius taken to be equal to one, but is kept here explicit for clarity. Next, let us consider a perturbation of the above set to the one with new coordinates $\bm{x}_{i}^{j}=({x}_{i}^{j},{y}_{i}^{j})$ satisfying the constraints analogous to \eqref{eqs:coordinate-constraints} and
\begin{align}
\label{eqs:coordinates-perturbations}
&x_{i}^{j-1}=\overline{x}_{i}^{j-1}+\xi_{i}^{j-1}, \quad
&y_{i}^{j-1}=\overline{y}_{i}^{j-1}+\eta_{i}^{j-1},
\end{align}
assuming the deviation $\bm{\xi}_{i}^{j-1}=(\xi_{i}^{j-1},\eta_{i}^{j-1})$ of the perturbed set from the initial one to be small compared to the ball radius, $|\bm{\xi}|\ll r$. The deviations $\bm{\xi}_i^j=({\xi}_i^j,{\eta}_i^j)$ in the linear order can be described by an effective continuous vector field $\bm{\xi}(\bm{x})$ defined on the unperturbed lattice $\overline{\bm{x}}_i^j$. After straightforward algebra of expanding the deviation field around its value at $\bm{\xi}_{i}^{j}$ we arrive at the system
\begin{subequations}
\label{eqs:hyperbolic}
\begin{align}
&\xi_{yy} -\left[\frac{1}{\beta\gamma}\,\xi_{x}\right]_x+\left[\left(\frac{1}{\beta}-\frac{1}{\gamma}\right)\xi_{x}\right]_y=0 ,\\
&\eta_{xx} -[\beta\gamma\,\eta_{y}]_y+[(\beta-\gamma)\eta_{y}]_x=0,
\end{align}
\end{subequations}
where $\beta(x,y)=\cot{\phi_{i}^{j}(x,y)}$ and $\gamma(x,y)=\cot{\psi_{i}^{j}(x,y)}$. Equations \eqref{eqs:hyperbolic} are of the hyperbolic type, i.e. one can interpret small perturbations as `propagating' along the characteristics in an inhomogeneous medium described by the functions $\beta(x,y)$ and $\gamma(x,y)$. In the case when $\overline{\bm{x}}_{i}^{j}=(\overline{x}_{i}^{j},\overline{y}_{i}^{j})$ is a regular pile, we have $\phi=\psi=\mbox{const}$, so that equations \eqref{eqs:hyperbolic} become \eqref{eqs:hyperbolic-regular}.

Next, for the $i$-th ball in the layer $j>0$, which is not on the slope, we get a set of equations
\begin{widetext}
\begin{subequations}
\label{balance:forces}
\begin{align}
\label{balance:force-x}
&\grave{N}_i^j\sin\phi_i^j-\grave{F}_i^j\cos\phi_i^j-\acute{N}_i^j\sin\psi_i^j+\acute{F}_i^j\cos\psi_i^j \nonumber \\
&\qquad+\acute{N}_{i-1}^{j+1}\sin\psi_{i-1}^{j+1}-\acute{F}_{i-1}^{j+1}\cos\psi_{i-1}^{j+1}-\grave{N}_i^{j+1}\sin\phi_i^{j+1}+\grave{F}_i^{j+1}\cos\phi_i^{j+1}=0, \\
\label{balance:force-y}
&\grave{N}_i^j\cos\phi_i^j+\grave{F}_i^j\sin\phi_i^j+\acute{N}_i^j\cos\psi_i^j+\acute{F}_i^j\sin\psi_i^j \nonumber \\
&\qquad-\acute{N}_{i-1}^{j+1}\cos\psi_{i-1}^{j+1}-\acute{F}_{i-1}^{j+1}\sin\psi_{i-1}^{j+1}-\grave{N}_i^{j+1}\cos\phi_i^{j+1}-\grave{F}_i^{j+1}\sin\phi_i^{j+1}-W=0, \\
\label{balance:torque}
&\grave{F}_i^j-\acute{F}_i^j-\acute{F}_{i-1}^{j+1}+\grave{F}_i^{j+1}=0,
\end{align}
\end{subequations}
\end{widetext}
with the notations for forces shown in fig.~\ref{fig:force-notations}. If the unperturbed irregular solution for forces is $\grave{\overline{N}}_i^j$, $\grave{\overline{F}}_i^j$, $\acute{\overline{N}}_i^j$, $\acute{\overline{F}}_i^j$ related with the same friction coefficient $\overline{\mu}=\mu_{1}$, then taking into account that angles and ball coordinates can expressed via
\begin{subequations}
\begin{align}
&\overline{x}_{i}^{j} - \overline{x}_{i}^{j-1}=2r\sin\overline{\phi}_i^{j}, &
&\overline{y}_{i}^{j} - \overline{y}_{i}^{j-1}=2r\cos\overline{\phi}_i^{j} \\
&\overline{\beta}=\cot{\overline{\phi}_{i}^{j}(x,y)}, &
&\overline{\gamma}=\cot{\overline{\psi}_{i}^{j}(x,y)},
\end{align}
\end{subequations}
we can determine from \eqref{balance:forces} the variations of forces $\delta \boldsymbol{N}_{i}^{j} = \left(\delta \grave{N}_i^j,\delta \acute{N}_i^j\right)$, $\delta \boldsymbol{F}_{i}^{j} = \left(\delta{\grave{F}}_i^j,\delta{\acute{F}}_i^j\right)$ as well as angles $\delta{\phi}_i^j$, $\delta{\psi}_i^j$. Proceeding as in the geometric problem and assuming, for simplicity, that all background angles equal $\phi_i^{j+1}=\phi_i^{j}=\psi_{i-1}^{j+1}=\psi_{i}^{j}=\phi$ (regular pile), as well as
\begin{align}
\label{variations:forces}
\delta \grave{F}_i^j=\mu_1 \delta \grave{N}_i^j, \quad \delta \acute{F}_i^j=\mu_2 \delta \acute{N}_i^j,
\end{align}
we arrive at the linear rheological relation $\boldsymbol{\sigma}(\nabla \boldsymbol{\xi})$ between the stress $\boldsymbol{\sigma}$ and deformation $\nabla \boldsymbol{\xi}$ tensors, which in the considered case simplifies to
\begin{subequations}
\label{relation:rheological-linear}
\begin{align}
&\grave{\sigma} B_{1} - \acute{\sigma} B_{2} =-\xi_y W, \\
\label{relation:rheological-linear:b}
&\grave{\sigma} A_{1} + \acute{\sigma} A_{2}= -\xi_x\left( W-2\frac{\grave{N}_i^j-\grave{N}_i^{j+1}}{\cos\phi}\right);
\end{align}
\end{subequations}
here $\grave{\sigma}=\delta \grave{N}_i^j-\delta \grave{N}_i^{j+1}$ and $\acute{\sigma}=\delta \acute{N}_i^j-\delta \acute{N}_{i-1}^{j+1}$ are the projections $\grave{\sigma} = - 2 \, r \, \grave{\boldsymbol{n}} \cdot \boldsymbol{\sigma} \cdot \grave{\boldsymbol{n}}$ and $\acute{\sigma} = - 2 \, r \, \acute{\boldsymbol{n}} \cdot \boldsymbol{\sigma} \cdot \acute{\boldsymbol{n}}$ of $\boldsymbol{\sigma}$ on the corresponding normal vectors $\grave{\boldsymbol{n}} = \grave{\boldsymbol{N}}_{i}^{j}/|\grave{\boldsymbol{N}}_{i}^{j}|$ and  $\acute{\boldsymbol{n}} = \acute{\boldsymbol{N}}_{i}^{j}/|\acute{\boldsymbol{N}}_{i}^{j}|$, $A_{i}=\cos\phi+\mu_{i}\sin\phi$, $B_{i}=\sin\phi-\mu_{i}\cos\phi$, and \eqref{relation:rheological-linear} is subject to the constraint due to torque balance
\begin{align}
\mu_1(\delta \grave{N}_i^j+\delta \grave{N}_i^{j+1})-\mu_2(\delta \acute{N}_i^j+\delta \acute{N}_{i-1}^{j+1})=0.
\end{align}
Note that since $\mu_{1,2}$ in \eqref{variations:forces} are considered known, then $\boldsymbol{\sigma}$ can be expressed in terms of variations of the normals forces only, i.e. $\boldsymbol{\sigma} = \nabla (\delta \boldsymbol{N})$. Since the determinant of the left-hand side of \eqref{relation:rheological-linear}
\begin{align}
\det=(1-\mu_1\mu_2)\sin\phi\cos\phi+(\mu_1+\mu_2)(\sin^2\phi-\cos^2\phi) \nonumber
\end{align}
may change its sign depending on the values of the friction coefficients $\mu_{1,2}$ and angle $\phi$, the characteristic type of the rheology changes between hyperbolic and elliptic. Given the expansion about node $(i,j)$:
\begin{subequations}
\begin{align}
\label{expansion:forces}
&\!\!\!\delta \grave{N}_i^{j+1}=\delta \grave{N}_i^{j}+2r [ \delta \grave{N}_x \sin{\phi}+\delta \grave{N}_y \cos{\phi}] , \\
&\!\!\!\delta \grave{F}_i^{j+1}=\delta \grave{F}_i^{j}+2r [\delta  \grave{F}_x \sin{\phi}+\delta \grave{F}_y \cos{\phi}] , \\
&\!\!\!\delta \acute{N}_{i-1}^{j+1}=\delta \acute{N}_{i}^{j}+2r [ -\delta \acute{N}_x \sin{\phi}+\delta \acute{N}_y \cos{\phi}] , \\
&\!\!\!\delta \acute{F}_{i-1}^{j+1}=\delta \acute{F}_{i}^{j}+2r [ -\delta \acute{F}_x \sin{\phi}+\delta \acute{F}_y \cos{\phi}] ,
\end{align}
\end{subequations}
and replacing $\grave{N}_i^j-\grave{N}_i^{j+1}$ on the rhs of \eqref{relation:rheological-linear:b} from \eqref{eqs:i1j} simplified under the introduced above conditions, we arrive at \eqref{eqn:rheology}. Notably, analogous perturbation expansion was attempted recently in a different context \cite{Acharya:2020,Acharya:2021}.

\vspace{0.5cm}

\noindent \textbf{Data availability}

The data that support the findings of this study are available from the authors upon reasonable request.\\

\noindent \textbf{Author contributions}

R.K. conceived the project and wrote the paper. A.Z. produced data. All authors
analyzed the results and commented on the manuscript.\\

\noindent \textbf{Competing interests}

The authors declare no competing interests.

%


\end{document}